A. G. Magner[1,2], A. I. Sanzhur[1], S. N. Fedotkin[1],
A. I. Levon[1], U. V. Grygoriev[1,3], S. Shlomo[2,*]

[1] *Institute for Nuclear Research, National Academy of Sciences of Ukraine, Kyiv, Ukraine*
[2] *Cyclotron Institute, Texas A&M University, College Station, Texas, USA*
[3] *Faculty of Science and Engineering, University of Groningen, Groningen, Netherlands*

*Corresponding author: s-shlomo@tamu.edu

# NUCLEAR LEVEL DENSITY
# IN THE STATISTICAL SEMICLASSICAL MICRO-MACROSCOPIC APPROACH

Level density $\rho$ is derived for a finite system with strongly interacting nucleons at a given energy $E$, neutron $N$ and proton $Z$ particle numbers, projection of the angular momentum $M$, and other integrals of motion, within the semiclassical periodic-orbit theory (POT) beyond the standard Fermi-gas saddle-point method. For large particle numbers, one obtains an analytical expression for the level density which is extended to low excitation energies $U$ in the statistical micro-macroscopic approach (MMA). The interparticle interaction averaged over particle numbers is taken into account in terms of the extended Thomas-Fermi component of the POT. The shell structure of spherical and deformed nuclei is taken into account in the level density by the Strutinsky shell correction method through the mean-field approach used near the Fermi energy surface. The MMA expressions for the level density $\rho$ reaches the well-known macroscopic Fermi-gas asymptote for large excitation energies $U$ and the finite combinatoric power-expansion limit for low energies $U$. We compare our MMA results for the averaged level density with the experimental data obtained from the known excitation energy spectra by using the sample method under statistical and plateau conditions. Fitting the MMA $\rho$ to these experimental data on the averaged level density by using only one free physical parameter – inverse level density parameter $K$ – for several nuclei and their long isotope chain at low excitation energies $U$, one obtains the results for $K$. These values of $K$ might be much larger than those deduced from neutron resonances. The shell, isotopic asymmetry, and pairing effects are significant for low excitation energies.

## 1. Introduction

Many properties of heavy nuclei were described in terms of the statistical level density [1-30]. For large excitation energies, in the statistical equilibrium, the level density has been derived in many works, starting from the Fermi gas formula derived in pioneer articles by Bethe [1] and Ericson [2],

$$\rho \propto \exp(S)/S^{\nu+1/2}, \quad \nu > 0, \tag{1}$$

where $S = 2\sqrt{aU}$ is the entropy, $a$ is the level density parameter, and $U$ is the excitation energy, and $\nu$ is related to the number of integrals of motion other than the total energy, see also Eq. (10) below. Obviously, the Fermi-gas formula, Eq. (1), diverges in the low energy limit, $U \to 0$. The first analytical attempt to remove this singularity in the low-excitation energies limit can be found in Ref. [31]. Notice that several other semi-analytical methods beyond the Fermi gas approach were suggested in Refs. [17,24,28,32-35]. The shell effects were studied in Ref. [31] in the yrast line energy as a function, $E_{\rm rot}(I)$, of the angular momentum, $I$, for "classical rotations" as an alignment of the individual angular momenta of particles along a symmetry axis in spherical nuclei within the periodic-orbit theory (POT) [36-42]. This semiclassical theory described the main shell structure of a finite Fermi system of strongly interacting Fermi particles, such as a nucleus, rather accurately. The main shell effects are related to the stability of nuclei with varying neutron or proton numbers, nuclear deformation, and spin. The nuclear energy can be presented as a sum of the smooth extended Tomas-Fermi (ETF) part [43-45] and its oscillating POT shell corrections of the Strutinsky shell correction method (SCM) [46,47]. Smooth and oscillating properties of the nuclear level-density parameter $a$ were first treated for neutron resonances; experimentally, e.g., in Refs. [5,21], and theoretically, within the main ETF approach in Refs. [44,45]. Then, the micro-macroscopic approach (MMA) was extended in Refs. [48-51] by working out the shell, isotopic asymmetry, and pairing effects. They were intensively worked out in Refs. [48-51], within the MMA approximation, for low excitation-energies spectra in spherical and deformed nuclei.

Why do we need the level density for low excitation energies along with high ones? For instance, it is often convenient to use the cumulative level density, $\mathcal{N} = \int_0^U dU' \rho(U')$, as a function of the excitation energy $U$, for analysis of the experimental data. For example, it can be applied for the study of the collective states, excited in the (p,t) two-neutron transfer reactions; see Ref. [52] and references therein. Another famous example is the fission width, $\Gamma_f = \int_0^{E-E_f} \rho(U) dU$, within the Bohr&Wheeler theory of nuclear fission, where $E_f$ is the fission barrier energy [53]. So, we have to unify analytically the micro-canonical (combinatoric formula for the level density $\rho$ at small excitation energies $U$ [54]) and macroscopic grand-canonical (Fermi gas formula (1) [2] for the level density $\rho$ at large excitation energies $U$) approaches within one statistical theory.

In the calculation of level density at low excitation energies in nuclei, we will consider the system of interacting Fermi particles with a macroscopic number $A = N + Z$, described by the Hamiltonian of the well-known nuclear superfluid model [55,56], taking the simplest Bardeen–Cooper–Schrieffer (BCS) [57] theory of the superconductivity [58]. The statistical characteristics of such a system can be calculated with the help of the mathematical apparatus of the superconductivity [57]. For this aim, one can transform the two-body Hamiltonian with the pairing interaction to that of the model Hamiltonian for independent quasiparticles by using the canonical Bogoliubov transformations [59]. This method was used also in the level density calculations [9,11], and for other purposes as nuclear astrophysics; see, e.g., Ref. [60]. We should emphasize the famous selfconsistent method of the superfluidity calculations named as the Hartree-Fock-Bogoliubov (HFB) theory [58,59]. See also its applications for calculations of ground state properties through the whole periodic table of nuclei (see Ref. [61]), that is in particular used in this work. Critical point for the superfluid-normal phase transition was studied in Refs. [62,63].

One of the most important phenomenon for nuclei, also important for nuclear superfluidity, is the shell structure, due to the inhomogeneity of the quasiparticle spectra near the Fermi surface energy, based on the Strutinsky shell correction method [46,47]. For a deeper understanding of the correspondence between the classical and the quantum approach, it is worthwhile to analyze the shell effects in the level density $\rho$, see Ref. [31], beyond the standard saddle point method (SPM) [2], within the semiclassical periodic-orbit (PO) theory (POT) [39,41,42,64]. Another root of applying the POT and SCM for the shell, isotopically asymmetry and pairing effects in the semiclassical level density calculations for spherical and deformed nuclei within the standard SPM was suggested in Ref. [15]. We extended the MMA approach [31] in Refs. [48-51] for the semiclassical description of these effects in nuclei in terms of the level density beyond the SPM. Smooth properties of the level density as function of the particle numbers have been studied within the framework of the self-consistent ETF approach [14,43,44]. However, for instance, the shell and pairing effects in the statistical MMA level density $\rho$ are still attractive subjects [9,11] for low excitation energies.

In the present work, we will present the POT shell-structure isotopically asymmetric MMA results for the level density $\rho(E,N,Z)$ taking into account paring correlations through the condensation energy shifts. We shall consider the influence of the pairing effects on the level density through the moments of inertia in a forthcoming work. This work is concentrated on the statistically averaged level density integrated over spins and their projections for low energy states in nuclear excitation-energy spectra below neutron resonances.

The structure of the paper is the following. The level density $\rho$ is derived within the MMA by using the semiclassical periodic-orbit theory in Section 2. The POT shell structure of the level density parameter $a$ is analyzed in Section 3. The basic final formulas used for the description of the main pairing effects are presented in Section 4. Section 5 is devoted to discussions of the results. The shell, pairing, and isotopic-asymmetry effects are discussed through several examples of nuclear spectra. Our results and perspectives are summarized in Section 6.

## 2. Level density derivations

We begin with a general micro-canonical definition of the level density:

$$\rho(E,\mathbf{Q}) = \sum_i \delta(E-E_i)\delta(\mathbf{Q}-\mathbf{Q}_i) \propto \int \lambda_E^3 \, d\lambda_E \, d\boldsymbol{\lambda} \, \exp[S(\lambda_E,\boldsymbol{\lambda})], \qquad (2)$$

where $E_i$ is the energy spectrum, $\mathbf{Q}_i$ is other quantum numbers for the nuclear state $i$, for instance, particle number $A_i$, or the number of neutrons $N_i$ and protons $Z_i$, projections of angular momentum, $M_i$, and so on. The Lagrange multipliers $\lambda_E$ and $\boldsymbol{\lambda}$ are related to the conservation of the energy $E$ and other integrals of motion $\mathbf{Q}$. Here, $S$ is the nuclear entropy,

$$S = \ln \mathcal{Z} + \lambda_E(E - \boldsymbol{\lambda}\mathbf{Q}), \quad \mathbf{Q} = \{N, Z, M\}, \quad \boldsymbol{\lambda} = \{\lambda_N, \lambda_Z, \lambda_M\}, \qquad (3)$$

where $\mathcal{Z}$ is the partition function,

$$\mathcal{Z} = \sum_i \exp[\lambda_E(E_i - \boldsymbol{\lambda}\mathbf{Q}_i)]. \qquad (4)$$

The inverse Laplace transform integrations in Eq. (2) are carried out along the straight line contour, parallel to the imaginary axis in the complex plane of a given variable, from the set of $\boldsymbol{\lambda}$, on the right of all poles of the partition function $\mathcal{Z}$. The partition function $\mathcal{Z}$ is sufficient for a full description of the statistical properties of any Fermi systems of strongly interacting particles. However, it is a very complicated quantity because for its calculation we have to know the quantum spectrum $E_i$ and $\mathbf{Q}_i$. Therefore, to simplify these calculations, it is especially very useful to derive the statistically averaged level density $\rho$ [Eq. (2)] analytically, beyond the Fermi gas model, solving the divergence problem at low excitation energies.

For the integrations in Eq. (2) we assume that it is possible to apply the saddle point method for all Lagrange multiplies, $\boldsymbol{\lambda}$, except for $\lambda_E$ which is related to the total energy $E$. Expanding the entropy $S$ near the saddle point $S^*$ over variables $\boldsymbol{\lambda}$, determined by the equation $\delta S = 0$, one has

$$S = S^* + \frac{1}{2}\left(\frac{\partial^2 S}{\partial N^2}\right)^*(N-N^*)^2 + \frac{1}{2}\left(\frac{\partial^2 S}{\partial Z^2}\right)^*(Z-Z^*)^2 + \frac{1}{2}\left(\frac{\partial^2 S}{\partial M^2}\right)^*(M-M^*)^2 + \dots, \qquad (5)$$

where $S^*$ is the entropy at the saddle point, $\boldsymbol{\lambda} = \boldsymbol{\lambda}^*$, $S^* = S(\lambda_E, \boldsymbol{\lambda}^*)$. For simplicity, one may omit nondiagonal contributions. Therefore, with the saddle point conditions, one has the conservation laws in the equations for the saddle point values $\boldsymbol{\lambda}^*$,

$$\mathbf{Q} = \lambda_E^{-1} \left( \frac{\partial \ln \mathcal{Z}}{\partial \boldsymbol{\lambda}} \right)^* \equiv -\left( \frac{\partial \Omega}{\partial \boldsymbol{\lambda}} \right)^*. \qquad (6)$$

Here, we introduced the generalized grand-canonical potential

$$\Omega \equiv -\ln \mathcal{Z} / \lambda_E = E_0 - a/\lambda_E^2 - \boldsymbol{\mu}\mathbf{N} - \Theta\omega^2/2, \qquad (7)$$

where $E_0$ is the ground state energy, $a$ is the level density parameter, $\boldsymbol{\mu} = \{\mu_n, \mu_p\} = \{\lambda_N, \lambda_Z\}$ stands for the neutron and proton chemical potentials, $\mathbf{N} = \{N, Z\}$, $\omega$ is the rotation frequency, and $\Theta$ is the moment of inertia of the rotating nucleus. The conservation laws (6) determine the Lagrange multipliers $\boldsymbol{\lambda}$ at the saddle point, $\boldsymbol{\lambda}^*$, in terms of the neutron $N$ and proton $Z$ numbers, and in terms of the projection of the angular momentum $M$. The coefficient, which appears in front of the exponent in the integrations, Eq. (2), over $\boldsymbol{\lambda}$ by the saddle point method, is the Jacobian:

$$J = \det\left( \frac{\partial \Omega}{\partial \boldsymbol{\lambda}}, \boldsymbol{\lambda} \right). \qquad (8)$$

For sufficiently low excitation energies $U$, for which one can nevertheless assume $T \ll \mu$ ($T \sim \sqrt{U/a}$, as explained in Refs. [48,49]), we derive simple analytical expressions for the level density $\rho(E, N, Z)$, beyond the standard SPM, by using accurate inverse Laplace integration (2) of the approximate integrand over $\lambda_E$. The Jacobian factor $J^{-1/2}$ in this integrand can be simplified much by expanding it in small values of $\xi$ or of $1/\xi$ (Ref. [48]), where

$$\xi \approx \frac{8\pi^6 U A^{1/3}}{3a\mu^2} \mathcal{E}_{sp}, \qquad \mathcal{E}_{sp} = -\frac{\delta E}{E_{ETF}} A. \qquad (9)$$

Here, $\mu \approx \mu_n \approx \mu_p$ is approximately the chemical potential near the nuclear stability line, and $\delta E$ is the relative energy shell correction modified by the pairing contributions ($\mathcal{E}_{sp}$ is associated with $E_{s+p}$, the calculated ground state shell-plus-pairing correction in notations of Ref. [61]). In Eq. (9), $E_{ETF}$ is the ETF energy component. In the applications below we will use $\xi > 0$ ($\mathcal{E}_{sp} > 0$). For $\mu = 40$ MeV, $A \sim 200$, and $\mathcal{E}_{sp} = |\delta E A/E_{ETF}| \approx 2.0$ [47,61], one finds the estimates $\xi \sim 0.1-10$ for temperatures $T \sim 0.1-1$ MeV. This corresponds approximately to a rather wide excitation energies $U = 0.2-20$ MeV for inverse level density parameter $K = A/a = 10$ MeV (Ref. [44]) ($U = 0.1-10$ MeV for $K = 20$ MeV). This energy range includes the low-energy states and states significantly above the neutron resonances. Within the POT [39,41,42] and ETF approach [41,43,45], these values are given finally by using the realistic smooth energy $E_{ETF}$ for which the binding energy [61] is $BE \sim E_{ETF} + \delta E$.

Expanding now the Jacobian factor, $J^{-1/2}$, at linear order in $\xi$ and $1/\xi$, one arrives at two different approximations marked below by cases (i) and (ii), respectively. Then, taking the accurate inverse Laplace transformation over $\lambda_E$ in Eq. (2), with the transformation of $\lambda_E$ to the inverse variable, $1/\lambda_E$, more accurately (beyond the standard SPM) [2], one approximately obtains (see Ref. [48]),

$$\rho \approx \rho_{\text{MMA}}(S) \approx \bar{\rho}_\nu f_\nu(S), \quad f_\nu(S) = S^{-\nu} I_\nu(S). \tag{10}$$

The argument $S$ of the modified Bessel functions, $I_\nu(S)$, of the order $\nu$ is the entropy $S$ given by $S = 2\sqrt{aU}$. For small, $\xi \ll 1$, case (i), and large, $\xi \gg 1$, case (ii), critical shell-structure quantity $\xi$ [$\xi \propto \mathcal{E}_{\text{sp}}$, Eq. (9)], one finds $\nu = 2$ and $3$, respectively. In cases (i) and (ii), called below the MMA1 and MMA2 approximations, respectively, one obtains Eq. (10) with different coefficients $\bar{\rho}_\nu$ (see Ref. [49]),

$$\rho_{\text{MMA1}}(S) \approx \bar{\rho}_2 S^{-2} I_2(S), \quad \bar{\rho}_2 \approx \frac{2\pi a}{3} \quad \text{(i)}, \tag{11}$$

$$\rho_{\text{MMA2}}(S) \approx \bar{\rho}_3 S^{-3} I_3(S), \quad \bar{\rho}_3 \approx \frac{4\pi a^2}{3\sqrt{\bar{\xi}}} \quad \text{(ii)}, \tag{12}$$

where $\bar{\xi} = 8\pi^6 A^{1/3} \mathcal{E}_{\text{sp}} / (3\mu^2)$. For the TF approximation to the coefficient $\bar{\rho}_3$ within the case (ii) ($g_{\text{ETF}} \propto A/\mu$), one finds [48,50]

$$\rho_{\text{MMA2b}}(S) \approx \bar{\rho}_3^{(2b)} S^{-3} I_3(S), \quad \bar{\rho}_3^{(2b)} \approx \frac{2\sqrt{6}\mu a^2}{3}. \tag{13}$$

In the derivation of the coefficient, $\bar{\rho}_3^{(2b)}$, we assume in Eq. (12) for $\bar{\rho}_3$ that the magnitude of the relative shell corrections $\mathcal{E}_{\text{sp}}$, $\xi \propto \mathcal{E}_{\text{sp}}$, see Eq. (9), can be negligibly small but their derivatives yield large contributions through the classical action of the oscillating POT level density derivatives $g''(\mu)$. For large entropy $S$, one finds from Eq. (10)

$$f(S) = \frac{\exp(S)}{S^\nu \sqrt{2\pi S}} \left[ 1 - \frac{1 - 4\nu^2}{8S} + \mathcal{O}\left(\frac{1}{S^2}\right) \right]. \tag{14}$$

At small entropy, $S \ll 1$, one obtains also from Eq. (10) the finite combinatorics power expansion [2,54]:

$$f(S) = \frac{2^{-\nu}}{\Gamma(\nu+1)} \left[ 1 + \frac{S^2}{4(\nu+1)} + \mathcal{O}(S^4) \right], \tag{15}$$

where $\Gamma(x)$ is the gamma function. This expansion over powers of $S^2 \propto U$ is the same as that of the phenomenological constant (effective) temperature (CT) model [3,27], used often for the level density calculations at small excitation energies $U$, but here, as in Ref. [48], we have it without free fitting parameters.

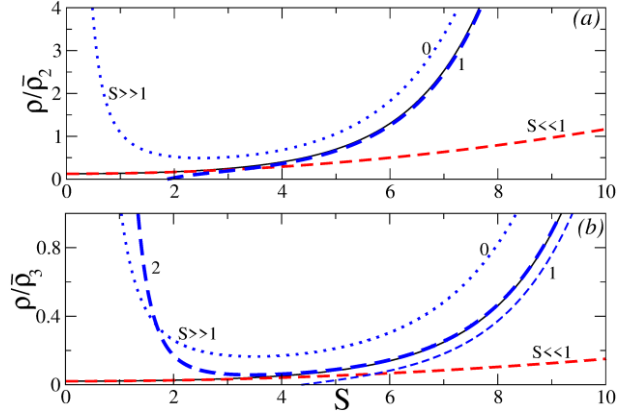

Fig. 1. Level density $\rho$, Eq. (10), in units of constant $\bar{\rho}_\nu$ as function of the entropy $S$, and its asymptotes, Eqs. (15) and (14); (a) and (b) for $\nu = 2$ and 3, respectively. "0" and "1" in (a) and "0", "1" and "2" in (b) show slow convergence of the contributions of the powers in $1/S$ of the expansion (14).

Fig. 1 shows the level density $\rho_{\text{MMA}}$, Eq. (10) (solid lines), in units of $\bar{\rho}_\nu$, as function of the entropy $S$ for different approximations. In Fig. 1 we present the level density dependence $\rho(S)$, Eq. (11) for $\nu = 2$ in (a), and Eq. (12) for $\nu = 3$ in (b), and their asymptotes. For small $S$ [red dashed lines, Eq. (15), "$S \ll 1$" in this figure] we present the combinatoric asymptote [2,54,65]. The only quadratic terms, $S^2 \propto U$, are taken into account to present similarly to the constant temperature model [3,27,28]. For large entropy $S$ [the asymptote $S \gg 1$] we neglected the corrections of the expansion in square brackets of Eq. (14), lines "0", "1" and "2" of the first (dotted), second [blue rare and frequent dashed lines in panels (a) and (b)], and third [rare dashed line in (b)] order terms over $1/S$ to show their slow convergence to the accurate MMA solid result, Eq. (10). It is interesting to find almost a constant shift of the leading asymptotic result at large $S$ (dotted line "0"), $\rho \propto \exp(S)/S^{\nu+1/2}$, in respect to the accurate MMA result of Eq. (10) (solid lines). This may clarify one of the phenomenological popular models – the back-shifted Fermi-gas (BSFG) model – for the level density calculations [3,11,21,66].

Using the standard SPM approach for calculations of the inverse Laplace integral over all variables, including $\lambda_E$, in Eq. (2), but keeping the shell and isotopic asymmetry effects (Ref. [49]), one arrives at

$$\rho(E, N, Z) \approx \frac{\sqrt{\pi} \exp\left(2\sqrt{aU}\right)}{12 a^{1/4} U^{5/4} \sqrt{1+\xi}}, \qquad (16)$$

where $\xi$ is given by Eq. (9) for a small asymmetry parameter, $X^2 = (N-Z)^2/A^2$, $\mu_n \approx \mu_p \approx \mu$. Eq. (16) is a more general shell-structure Fermi-gas (SFG) asymptote, at large excitation energy, with respect to the well-known [1,2] Fermi gas (FG) approximation for $\rho(E,N,Z)$, which is equal to Eq. (16) at $\xi \to 0$,

$$\rho(E,N,Z) \to \frac{\sqrt{\pi}\exp\left(2\sqrt{aU}\right)}{12a^{1/4}U^{5/4}}. \qquad (17)$$

Notice that in the derivations of the SFG and FG expressions we did not use the small spin approximation. Moreover, these expressions were obtained by integrating over all spin projections $M$. Therefore, Eqs. (16) and (17) do not contain the free cut-off parameter (proportional to the moment of inertia and temperature), in contrast to those used, e.g., in Refs. [18-21]. In contrast to the finite MMA limit (15) for the level density, Eq. (10), the asymptotic SFG [Eq. (16)] and FG [Eq. (17)] expressions are obviously divergent at $U \to 0$. Eqs. (16) and (17) can be obtained from the leading term of the asymptote (14) for large excitation energies $U$. Notice also that the MMA1 approximation for the level density, $\rho(E,N,Z)$, Eq. (11), can be applied also for large excitation energies, $U$, with respect to the collective rotational excitations, as the SFG and FG approximations if one can neglect shell effects, $\xi \ll 1$. Thus, the level density $\rho(E,N,Z)$ in the case (i), Eq. (11), has wider range of the applicability over the excitation energy variable $U$ than the MMA2 case (ii). The MMA2 approach has, however, another advantage of describing the important shell structure effects.

In Eq. (10), the argument $S$ of the modified Bessel functions of the index $\nu$, $I_\nu(S)$, depends on the level density parameter $a$, and the excitation energy $U$. In the Fermi gas model for large excitation energy $U$, the saddle point value of $\lambda_E$ is $\lambda_E^* = 1/T$, where $T$ is the temperature. Then, we arrive at the well-known excitation energy expression

$$U = E - E_0 - \Theta\omega^2/2. \qquad (18)$$

For small shell-structure contribution, one finds the index $\nu = \kappa/2 + 1$, where $\kappa$ is the number of additional integrals of motion beyond the energy $E$. This integer number is the dimension of $\mathbf{Q}$, $\mathbf{Q} = \{N, Z, ...\}$ for the case of two-component atomic nuclei, where $N$ and $Z$ are the numbers of neutrons and protons, respectively. For much larger shell structure contributions, one obtains a larger value of $\nu$, $\nu = \kappa/2 + 2$.

## 3. Level density parameter and POT shell effects

In Eq. (10) for the entropy $S$, $a$ (for a given isotopic index $\tau = \{n, p\}$, specifying neutron or proton component) is the level density parameter with the SCM decomposition:

$$a = \frac{\pi^2}{6} g_{scl}, \qquad g_{scl} = g_{ETF}(\mu) + \delta g_{scl}(\mu). \tag{19}$$

Here, $g_{ETF}$ is the sum of the neutron and proton terms of the extended Thomas-Fermi component and $\delta g_{scl}$ is the corresponding periodic-orbit shell correction. Therefore, $a$ is independent of the excitation energy $U$. Simple explicit expressions for the level density parameter, Eq. (19), with the $\omega$ dependence and its ETF and POT components in the case of a spherical mean field are given in Ref. [48]. For the extended Thomas-Fermi component [41,43-45], $g_{ETF}$, one takes into account the self-consistency by employing the Skyrme forces [45,67]. For the semiclassical PO level-density shell corrections, $\delta g_{scl}(\mu)$ (labeled by $\tau$, e.g., $\mu = \mu_\tau$), one can use the POT [38,39,41,42,48,64]. A mean (nuclear) potential is needed to specify these shell corrections in terms of the periodic orbits. For simplicity, we omit everywhere in this section the isotopic neutron-proton index $\tau$ [49].

The excitation energy $U$ for the entropy $S$ in Eqs. (10), and (18) depends explicitly on the shell effects,

$$U = E - (E_{ETF} + \delta E) - \frac{1}{2}(\Theta_{ETF} + \delta\Theta)\omega^2, \tag{20}$$

where $E_{ETF}$ and $\delta E$ are the smooth ETF and shell correction components of the background energy, $E_0 \approx E_{ETF} + \delta E$. Similar decomposition takes place for the moment of inertia, $\Theta = \Theta_{ETF} + \delta\Theta$. For the energy shell correction, $\delta E$, and the level density shell correction, $\delta g_{scl}$, one has the semiclassical expressions in terms of the sum over periodic orbits (POs),

$$\delta E \cong \sum_{PO} \left(\hbar^2 / t_{PO}^2\right) g_{PO}(\mu), \tag{21}$$

with,

$$\delta g_{scl}(\varepsilon) \cong \sum_{PO} g_{PO}(\varepsilon). \tag{22}$$

Here, $t_{PO}$ is the period of the particle motion along the PO at the single-particle (s.p.) energy $\varepsilon = \mu$, $t_{PO} = (\partial S_{PO} / \partial \varepsilon)_{\varepsilon=\mu}$, where $S_{PO}$ is the classical action along the PO,

$$g_{PO}(\varepsilon) = \mathcal{A}_{PO}(\varepsilon) \cos[S_{PO}(\varepsilon)/\hbar - (\pi/2)\mu_{PO} - \phi]. \tag{23}$$

In Eq. (23), the Maslov index $\mu_{PO}$ is determined by the integer number of the catastrophe (caustic and turning) points along the PO, and $\phi$ is a constant independent of PO but depends on the dimension of the problem and degeneracies of the PO families; see Refs. [38,39,41,42] for details and a lot of examples of the spherical and deformed potentials. As in the SCM, the POT level-density shell corrections sum (22), $\delta g_{\text{scl}}(\varepsilon)$, labeled by the isotope index $\tau = \{n, p\}$, is convergent due to the coarse-grained averaging over the s.p. energies of the corresponding potential well. The Gaussian local averaging of the level density shell correction, Eqs. (22) and (23), and therefore, the level density parameter $a$ through Eq. (19), is carried out over the s.p. energy $\varepsilon$ near the Fermi surface, $\mu \approx \varepsilon_F$, with a width parameter $\Gamma$. Taking $\Gamma$ smaller than a distance between major shells, $\mathcal{D}_{\text{sh}}$, this averaging can be done analytically with a good accuracy [39,41,42],

$$\delta g_{\text{scl},\Gamma}(\varepsilon) \cong \sum_{\text{PO}} g_{\text{PO}}(\varepsilon) \exp\left[-\left(\frac{\Gamma t_{\text{PO}}}{2\hbar^2}\right)^2\right]. \tag{24}$$

We should emphasize, however, that for the semiclassical energy shell correction $\delta E$, Eq. (21), and that of the moment of inertia, $\delta\Theta$, and of the corresponding thermodynamical potential, $\delta\Omega$, such an additional averaging is not needed because of the additional convergence factor, $1/t_{\text{PO}}^2$, which already guarantees a fast convergence over the PO length. The energy shell correction, $\delta E$, in Eq. (9) can be approximated, for a major shell structure, with the semiclassical POT accuracy (see Eq. (21) and Refs. [39,41,42,64]) by

$$\delta E \approx \delta E_{\text{scl}} \approx \left(\frac{\mathcal{D}_{\text{sh}}}{2\pi}\right)^2 \delta g_{\text{scl}}(\mu). \tag{25}$$

Here, $\mathcal{D}_{\text{sh}} \approx \mu / A^{1/3}$ is the distance between major shells near the Fermi surface energy of the order of the chemical potential $\mu$, and $\delta g_{\text{scl}}(\varepsilon)$ is the semiclassical single-particle level-density shell correction, taken at the energy $\varepsilon \approx \mu$, and oscillating function of the classical action over $\hbar$ [39,41]. The characteristic parameter $\xi$, Eq. (9), proportional to $\mathcal{E}_{\text{sp}}$, specifies the two different approximations, $\xi \ll 1$ and $\xi \gg 1$, for small and large shell correction contribution $\mathcal{E}_{\text{sp}}$, respectively. We will consider below magic nuclei, for which one has the case of relatively large $\mathcal{E}_{\text{sp}}$ and, therefore, large $\xi$.

For the semiclassical grand-canonical potential $\Omega_{\text{scl}}$, one has a similar decomposition, $\Omega_{\text{scl}} = \Omega_{\text{ETF}} + \delta\Omega_{\text{scl}}$. The oscillating semiclassical shell component, $\delta\Omega_{\text{scl}}$, of the grand-canonical potential $\Omega_{\text{scl}}$, is given at the saddle point $\lambda_E^* = 1/T$, with the temperature $T$, in the small rotation-frequency (adiabatic) approximation by the following PO sum [48,51]:

$$\delta\Omega_{\text{scl}}(T,\mu,\omega) \cong \sum_{\text{PO}} g_{\text{PO}}(\mu) \frac{\hbar^2}{t_{\text{PO}}^2(\mu)} \frac{\pi T t_{\text{PO}}/\hbar}{\sinh(\pi T t_{\text{PO}}/\hbar)} + \frac{1}{2}\delta\Theta(\mu)\,\omega^2, \tag{26}$$

and similarly, for the moment of inertia [48,51,68]. In the limit to zero temperature $T$, Eq. (26) converges to the energy shell correction $\delta E$ of Eq. (21). Besides of the same factor $1/t_{PO}^2$, one has the additional exponential temperature-dependent convergence factor in the PO sum (26). For large temperatures, $T \gg T_{sh}$, one finds an exponential decrease of $\delta \Omega_{scl}$ with a characteristic temperature for large particle numbers $A \approx 100-200$,

$$T_{sh} \approx \mathcal{D}_{sh} / \pi \approx 2-3 \, \text{MeV} . \tag{27}$$

We neglect here the isotopic asymmetry for $X^2 \ll 1$, $\mu \approx \mu_n \approx \mu_p$, for large particle numbers, $\mathcal{S}_{PO}(\mu)/\hbar \gg 1$. The disappearance of the shell correction at temperatures $T \approx T_{sh}$, Eq. (27), was estimated in terms of the distance $\mathcal{D}_{sh}$ between major shells:

$$\mathcal{D}_{sh} \approx \frac{2\pi\hbar}{\langle t_{PO} \rangle} \approx \frac{\mu}{A^{1/3}} = 7-10 \, \text{MeV} , \tag{28}$$

where $\langle t_{PO} \rangle$ is the average of the most important short and degenerate POs near the Fermi energy surface, $\varepsilon \approx \mu$. The temperature dependence of the level density parameter $a$ for large temperatures was discussed also in Refs. [12,13].

### 4. Pairing collapse and condensation energy

The pairing contributions to the nuclear level density and their collapse at a critical temperature have been discussed quite long time; see, e.g., Refs. [2,9-11,15,16,63,66]. We use the traditional method for taking into account the pairing effects in the level density [9,11] through the condensation energy shifts of the excitation energy $U$ in our semiclassical MMA approach [48] for low excitation energies. For simple solutions of the gap equation within the simplest BCS approach, the statistically averaged condensation energy $E_{cond}$ is derived in terms of the constant pairing gap $\Delta$, independent of the quasiparticle spectrum [58]. For the statistically averaged MMA level density $\rho(E, N, Z)$ up to particle number fluctuations, one can use its averaged empiric dependence on the particle number $A$, as $\Delta \approx \Delta_0 \approx 12 A^{-1/2}$ MeV [4,9,11,58]. The results for the level density are smoothly dependent on the factor in front of $A^{-1/2}$. This phenomenological behavior $\Delta(A)$ is rather good for sufficiently heavy nuclei, for $A \gtrsim 100$.

Following Refs. [9,11], one can reduce the level density calculation for the system of interacting Fermi-particles described by the two-body Hamiltonian to that of the system of a mean field with the quasiparticles' BCS Hamiltonian $\hat{H}_{BCS}$ [58]. The corresponding thermodynamical averages of any

operator $\hat{Q}$ are determined by

$$\langle\hat{Q}\rangle = \text{Tr}\left[\hat{Q}\exp(-\lambda_E \hat{H}_{\text{BCS}})\right] / \text{Tr}\left[\exp(-\lambda_E \hat{H}_{\text{BCS}})\right] \tag{29}$$

Up to a constant, the Hamiltonian $\hat{H}_{\text{BCS}}$ coincides with that of the Fermi-quasiparticles in a mean field. Therefore, for the entropy $S$, one can use a similar expression:

$$S_{\text{pair}} = 2\sum_i [\lambda_E \epsilon_i \bar{n}_i - \ln(1-\bar{n}_i)], \tag{30}$$

where $\epsilon_i = [(\varepsilon_i - \mu)^2 + \Delta^2]^{1/2}$, $\varepsilon_i$, and $\bar{n}_i = [1+\exp(\lambda_E \epsilon_i)]^{-1}$ are the quasiparticle energies, single-particle energies, and occupation number averages, respectively [9].

For relatively low excitation energies, $\Delta/T \gg 1$, but sufficiently large for using the standard SPM, and therefore, defining the temperature $T$ as the saddle point, $T = 1/\lambda_E^*$, the excitation energy dependence of the pairing gap is given by [9,11]

$$\Delta - \Delta_0 = -\sqrt{2\pi\Delta_0/\lambda_E}\exp(-\lambda_E \Delta_0). \tag{31}$$

At the saddle point, $\lambda_E = \lambda_E^* = 1/T$, one thus has the temperature behavior of the gap $\Delta(T)$. In the opposite asymptotic regime of relatively large excitation energies $U$, when the temperature $T$ can be determined as a saddle point, and when $T$ is close to the critical temperature $T_c$ for a destruction of pairing correlations, $T \to T_c$, the gap equation for $\Delta$ can be linearized by expanding it in power series over $\Delta/T$. Straightforward derivations [9,11] lead to the critical temperature $T_c$ found from Eq. (31) at the condition $\Delta(T) = 0$,

$$T_c = e^C \Delta_0 / \pi, \tag{32}$$

where $C \approx 0.577$ is the Euler constant.

For a given temperature $T$, when exists, by minimizations of the expectation value of the grand-canonical potential $\Omega$, one has (see Refs. [9,11]),

$$\Omega = \langle \hat{H}_{\text{BCS}} - \boldsymbol{\mu}\hat{\mathbf{N}} - \hat{S}/\lambda_E \rangle, \tag{33}$$

where $\langle \ldots \rangle$ denotes a statistical average (29) over the operator enclosed in angle brackets, Eq. (29). In Eq. (33), $\hat{\mathbf{N}}$ is the particle (neutron and proton) number and $\hat{S}$ is the entropy operators, which correspond to the above introduced neutron $N$, proton $Z$ and entropy $S_{\text{pair}}$ in a nucleus. For the pairing ground-state energy $\langle H_0 \rangle$ ($\langle H_{\text{BCS}} \rangle$ at zero excitation energy, $U = 0$), one finds $\langle H_0 \rangle \approx \Delta_0^2 / 4G$, where $G$ is the average estimate for the two-body interaction matrix elements [58]. With the heat part, $U_c = aT_c^2$, where $a$ is the level density parameter, for the total excitation energy

$U_{\text{c}}^{\text{tot}}$ at the critical temperature $T_{\text{c}}$ [Eq. (32)] for the pairing superfluid-normal phase transition, one obtains

$$U_{\text{c}}^{\text{tot}} = aT_{\text{c}}^2 + \Delta_0^2/4G .  \qquad (34)$$

Notice that a sharp pairing collapse at the critical temperature $T_{\text{c}}$, Eq. (32), takes place for the nuclear systems described by the mean field BCS or HFB approximation, neglecting the statistical fluctuations, e.g., the particle number fluctuations [1], which can be estimated roughly as $\left(\langle A^2 \rangle - \langle A \rangle^2\right)^{1/2} \sim 1-2$. As shown in Refs. [62,63], more realistic calculations of the pairing potential $\Delta$ by the HFB mean field with accounting for the particle number fluctuations lead to a blur of the pairing superfluid-normal phase transition over temperature, $\Delta(T)$, in heavy finite nuclei. Similar blur takes place for the statistical averaging of the pairing potential $\Delta$. Such a phenomenon is in disagreement with strong statements on absence of the pairing effects in the magic close-shell nuclei. However, as shown below, due to the statistical averaging up to particle number fluctuations, this transition is not easily observed, for instance, in the nucleus $^{208}$Pb.

Accounting for the condensation energy $E_{\text{cond}}$ in derivations of Section 2, one can obtain simple analytical expressions for the level density $\rho(E,N,Z)$, beyond the standard SPM, but with pairing effects. According to the inverse Laplace transformation, one finds the expression (10) for the level density as function of the entropy, $S \to S_{\text{pair}}$, given by

$$S_{\text{pair}} = 2\sqrt{aU_{\text{eff}}} .  \qquad (35)$$

The level density parameter $a$ is independent of the excitation energy, $U_{\text{eff}}$, shifted due to the pairing correlations below the critical excitation energy $U_{\text{c}}^{\text{tot}}$, Eq. (34),

$$U_{\text{eff}} = U - E_{\text{cond}} .  \qquad (36)$$

For the condensation energy $E_{\text{cond}}$, one finally has (see Refs. [9,11])

$$E_{\text{cond}} = \frac{3a\Delta_0^2}{2\pi^2}  \qquad (37)$$

$$\approx \frac{1}{4} g_{\text{TF}} \Delta_0^2 \approx \frac{216}{\pi^2 K} ,  \qquad (38)$$

---

[1] The particle number fluctuations, $A_{\text{fl}} = \left(\langle A^2 \rangle - \langle A \rangle^2\right)^{1/2}$, can be evaluated using the classical Landau theory [7] as $A_{\text{fl}}^2 = -T\partial^2\Omega/\partial\mu^2 \sim (aU)^{1/4}$ ($a = A/K$). For large particle numbers $A \approx 100-200$, excitation energies $U \approx 1-2$ MeV in the low energy range, and typical inverse level-density parameters $K \approx 10-30$ MeV [44,49], one finds particle number fluctuation of $A_{\text{fl}} \approx 1-2$.

where $g_{TF}$ is the TF level density estimate, $g_{TF} \approx 3A/2\mu$, and $K = A/a$ is the inverse level density parameter. For $K \approx 10-30$ MeV [44,48,49], one obtains $E_{cond} \approx 1-2$ MeV. Notice also, that the condensation energy $E_{cond}$, Eq. (37), depends on the particle number $A$ through the inverse level density parameter $K$, Eq. (38), which is a slowly oscillating function of $A$ [48,49]. Again, for large and small $S$, from the general Eq. (10), shifted now by the condensation energy (37), one obtains the famous Fermi gas [2] and combinatoric [54] expressions with the important $1/S$ and $S^2$ corrections, respectively. In the pairing modified equation (10) with $S \to S_{pair}$ the value of $\nu$ depends also on the number of the integrals of motion and on the shell structure contribution, see Section 2. This contribution is measured by the parameter $\xi$, Eq. (9) but also with replacement $U \to U_{eff}$.

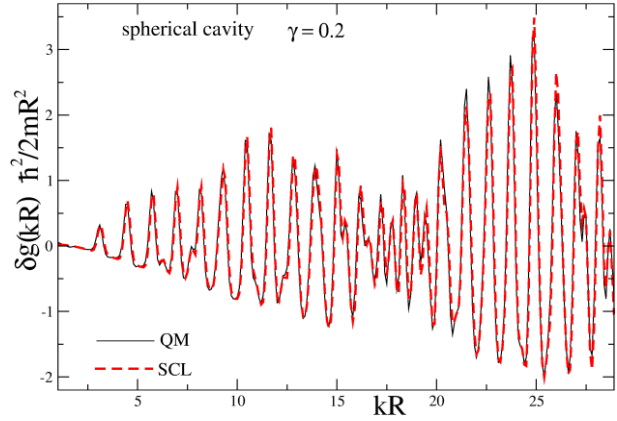

Fig. 2. The POT and SCM level density shell corrections $\delta g$ as functions of the dimensionless "energies" $kR$, where $k$ is the wave number, $k = \sqrt{2m\varepsilon/\hbar^2}$, $m$ is the nucleon mass, $R$ is the radius of the spherical cavity; see the details in the text.

## 5. Discussions of the results

In Fig. 2 we show the level density shell corrections for the spherical cavity. Solid and dashed lines present the results of quantum-mechanical (QM) and semiclassical (SCL) approaches, respectively. The POT level density shell corrections $\delta g$ were approximated by the famous analytical POT trace formula [38] for the infinitely deep spherical square-well potential. It is convenient to consider the dimensionless quantum energy spectrum, $k(\varepsilon)R$, where $k$ is the wave number and $R$ is the radius of the spherical cavity. For Strutinsky's smoothing procedure, a relatively large Gaussian width $\tilde{\gamma} = 2.0$, and a correction polynomial degree $\mathcal{M} = 6$, are used to satisfy the plateau condition [47]. With these parameters, $\tilde{\gamma}$ and $\mathcal{M}$, the result of the SCM calculations does not depend on their values within rather a large plateau range of these average parameters. For a small average Gaussian-width parameter, $\gamma \ll \tilde{\gamma}$, to keep the major shells (also, subshell, and supershell) structure we take $\gamma = 0.2$ for quantum states of the dimensionless $k(\varepsilon)R$ spectrum. This parameter is free for the POT level density correction averaging but common for all nuclear numbers in Fig. 3, and correspond to a major shell and subshell structure [41,42]. The corresponding dimensional Gauss width is $\Gamma = 2\gamma\mu/(k_F r_0 A^{1/3}) \approx 2$ MeV for $r_0 = 1.14$ fm and particle numbers $A \approx 100-200$. Notice that a similar result but for the major shell and supershell structure with $\gamma = 0.3$ was presented in Ref. [50]. The number of vertexes $u$ and winding number $w$ for leading POs contribution into the sum (17) near the chemical potential $\mu$ are restricted by the maximal winding number for periodic orbits (perfect polygons and diameters) in the spherical cavity, with $w_{max} = 10$ (in Ref. [50] for $\gamma = 0.3$, the dominating winding number, $w_{max}$, is much smaller). The chemical potential $\mu$ for heavy nuclei is rather constant, $\mu = 40$ MeV, independent of particle number $A$. As seen from Fig. 2, there is almost no difference between the QM and SCL level density results for a given much smaller Gaussian parameter, $\gamma \ll \tilde{\gamma}$ as for $\gamma = 0.3$ in Ref. [50]. The nuclear region is rather small in this plot, $kR \lesssim 6$, and high values of $kR$ are related, e.g., to metallic clusters.

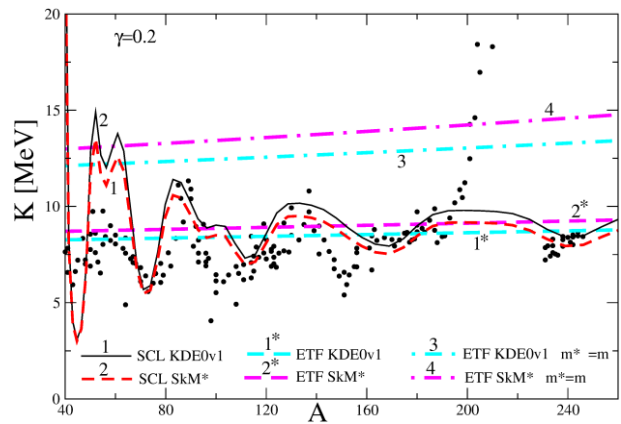

Fig. 3. The inverse level density parameter $K = A/a$ as a function of the particle number $A$ with the level density parameter $a$ (19), related to Fig. 2 through the total POT level density $g_{scl}$, for different Skyrme forces, KDE0v1 "1" [67] and SkM* "2"[43]. In the self-consistent ETF parts of these calculations [42], the rare dashed lines "1*" (or "2*") with the effective mass $m^*$, and dashed-dotted lines "3" (or "4") for $m^* = m$ are related to the same Skyrme forces KDE0v1 (or SkM*), respectively. Black dots correspond to the constrained experimental data taken from Ref. [18]; see the text and Ref. [50].

In Fig. 3 the results of calculations of the inverse level density parameter $K$ are compared with the values obtained in Ref. [18] from the analysis of constrained experimental data within the BSFG and

CT models. The shell correction density components, $\delta g_{\text{scl}}$, of Fig. 2 are used for more realistic calculations with taking into account the ETF parts $g_{\text{ETF}}$ of the POT level density $g_{\text{scl}}$, Eq. (19). One obtains the inverse level-density parameter, $K = A/a(\mu)$, as a function of the particle number $A$ (Fig. 3) by using the total (ETF and shell corrections) POT level density, $g_{\text{scl}}(\varepsilon)$, through Eq. (19) for the level density parameter $a$, at the chemical potential, $\varepsilon = \mu$. We emphasize that in all our calculations (Figs. 3-7) the inverse level density parameter $K$ is independent of the excitation energy $U$ (and $U_{\text{eff}}$) as it was assumed in the derivations of Eqs. (10)-(13), (16), and (17), in contrast to that assumed in Ref. [6]. The smooth part, $g_{\text{ETF}}$, of the total POT level density, $g_{\text{scl}}$, is approximated by the self-consistent extended Thomas-Fermi (ETF) approach for two versions of Skyrme forces [44,45]. One of them is the SKM* [43] (rare dashed line "2*" with accounting for the effective mass and rare dashed-dotted line "4" for effective mass $m^* = m$). Another one, shifted a little down to the results of the constrained experimental data [18], is the other KDE0v1 [67] Skyrme force; see rare dashed line "1*" and dotted-dashed line "3". Solid line "1" and frequent dashed line "2" oscillating curves show the corresponding values of inverse level density parameter $K$ due to the semiclassical (SCL) POT level-density contribution, $g_{\text{scl}} = g_{\text{ETF}} + \delta g_{\text{scl}}$, to the level density parameter $a$ [see Eq. (19), $a_{\text{scl}} = \pi^2 g_{\text{scl}}(\mu)/6$] with $\delta g_{\text{scl}}$ shown in Fig. 2. Other parameters are the same as in the previous Fig. 2. For simplicity, the effective mass contribution into the shell corrections can be neglected with respect to that of the ETF parts. The result of these calculations are largely in a qualitative agreement with the recent analysis of experimental data (Ref. [18]), which, as compared to Ref. [5], included in the analysis of many other excited nuclei and different reactions with nuclear excitation energies being significantly smaller than the neutron separation energy. The sets with reliable completeness of levels in the limited energy range below the neutron binding energy were selected for each nucleus in Ref. [18]. The neutron resonance level densities were also included in the analysis of Ref. [18], in contrast to our results. However, the weighted Least Mean-Square (LMS) fitting with large widths of the levels, used in the model-dependent calculations of the level-density parameter $a$ in Ref. [18], suppresses the contribution of the low energy states. Therefore, they are remarkably similar to the neutron resonance behavior presented earlier in Ref. [5].

In spite of obvious discrepancies of the curves $K(A)$, especially near the $^{208}$Pb, the major period of the nuclear shell structure is clearly recognized in Fig. 3 by the maxima of $K(A)$, which correspond to the minima of the level density parameter $a$, or approximately of the oscillating level density component. The relationship between the chemical potential $\mu$, through the Fermi momentum in $\hbar$ units is given by $k_F = \sqrt{2m\mu/\hbar^2}$, where $m$ is the nucleon mass, and $k_F R \propto A^{1/3}$, according to the relationship (6). Mean value of oscillating $K(A)$ in Fig. 3 is about 8 MeV, as predicted in Ref. [14]. This is in accordance with the ETF (SkM$^*$, or KDE0v1) value, associated with the effective mass $m^*$. As shown in Ref. [44], the effect of the effective mass $m^*$ on the inverse level density parameter $K$ is very strong. This leads approximately to mean values of the constrained experimental data [18], especially well for the value associated with the Skyrme force KDE0v1; see Fig. 3. However, we should not expect such good agreement with these experimental data using the level density shell corrections obtained for the infinitely deep spherical square-well potential without accounting for the spin-orbit interaction term in the oscillating component $\delta g_{scl}$. The positions of the minima (maxima) of $a$, i.e., of the single-particle level density $\delta g_\gamma(\varepsilon)$ at $\varepsilon = \mu$, related to magic nuclei in that potential, cannot be correctly reproduced in such shell-correction calculations, because of neglecting the spin-orbit interaction. In order to show the main effective-mass effect through the mean value of $K$, one may neglect its influence on $K$ through the oscillating part of the level density parameter $a$, Eq. (19). As shown in Refs. [42,64], in order to reproduce the experimental value of $K$ for the second minimum in the double-humped deformation-energy well, related to the intermediate state in the deformed nucleus $^{240}$Pu, in the quantum and semiclassical calculations, we should shift the curves $K(A)$ along the $A$ axis (through $k_F$). Therefore, we shifted the semiclassical curves in Fig. 3 with about $\Delta A = 20$ along the particle number $A$ axis. This shift is of the order of the period $A_{sh} \approx \mathcal{D}_{sh} g_{TF} \sim A^{2/3}$ ($g_{TF} \sim A/\mu$) related to the distance $\mathcal{D}_{sh}$, Eq. (28), between major shells near the Fermi surface energy. This is similar to the discussions in Ref. [42] where the intermediate state [47] in the deformed nucleus $^{240}$Pu was obtained semiclassically by using a similar shift. Therefore, three minima of the level density shell corrections $\delta g_{scl}$ for the major shell closures in the semiclassical calculations at $A \approx 45-150$ shown in Fig. 3 correspond to the maxima of the inverse level density parameter, $K = A/a$, obtained in Ref. [18] by using the constrained experimental data. In spite of very simple explicitly given analytical formulas of Refs. [38,39,42,64] for the POT shell corrections in the spherical cavity, one obtains a largely good agreement of the value of the semiclassical approximation for $K$ with the constrained experimental data [18]. The magnitudes of periods for the oscillations of $K(A)$ are basically in good agreement with experimental data in the range of particle numbers $A \approx 100-150$. The POT evaluations of the period of the shell structure $\mathcal{D}_{sh}$, Eq. (28), does not depend much on the nuclear deformation, and give approximately the correct estimation. This is in contrast to the amplitudes of oscillations of the curve $K(A)$ because, according to Eq. (28), it depends on the averaging Gaussian parameter $\gamma$ for the POT level density $\delta g_{\rm scl}$. Notice that even the constraint selfconsistent HF (or HFB) calculations with realistic Skyrme forces have the same problem with averaging of the s.p. level density shell correction [47]. However, there is a discrepancy between experimental and theoretical results for $K(A)$ in the range of particle numbers $A \approx 150-240$, and also for small particle numbers, $A \sim 40$. As mentioned above, the experimental data for $K$ are in good agreement with those of neutron resonances (see Refs. [5,44]) which are dominating in the results of the calculations of Ref. [18] using the experimental spectra data. We will show below that it is important to find $K$ from the LMS fitting of the theoretical results for the statistically averaged level density $\rho$ with that obtained from the experimental excitation-energy spectra by using its corresponding averaging procedure. In particular, to obtain the model-independent experimental results we study the plateau condition, as done in the shell correction method [47]. The specific reason for the discrepancy might be that the level density parameter $a$ (or $K$) was obtained by three free parameter fit of these experimental data to the level density within the BSFG and CT models. Another reason discussed in Section 4 is that the pairing effects should be taken into account properly below the critical excitation energy $U_{\rm c}^{\rm tot}$ for the pairing collapse.

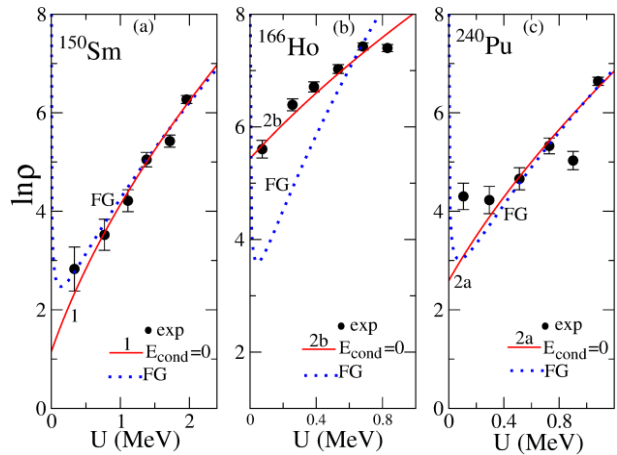

Fig. 4. The MMA level density $\rho$ (in logarithmic scale) and its FG asymptote as functions of the excitation energies $U$ and experimental data for average level density obtained by the sample method from the spectrum of the database [69]; see Table 1 and text.

Figs. 4, 5 and 6 show the level density, $\ln\rho(E, N, Z)$, Eq.(10), for low energy states in nuclei within our MMA approach for the smallest standard critical-error parameter of the LMS Fit (LMSF) $\sigma^2$,

$$\sigma^2 = \frac{\chi^2}{\aleph - 1}, \qquad \chi^2 = \sum_i \frac{(y(U_i) - y_i^{\exp})^2}{(\Delta y_i)^2}, \qquad y = \ln \rho. \qquad (39)$$

The important constants, and results of the LMS fitting of only one free parameter $K(\pm \Delta K)$, with corresponding pairing characteristics $E_{\text{cond}}$ and $U_c^{\text{tot}}$ in cases of notable values of pairing condensation energies, are obtained from these calculations and presented in Table 1. Experimental dots are obtained from the excited states spectrum of these nuclei (ENSDF database, Ref. [69]) by using the sample method. Energy spectrum states $E_i, I_i$ are distributed in sample boxes of the same energetic length, $U_{\text{box}}$. The numbers of states $L_i$ (including $2I_i + 1$ spin degeneracy) in the $i$ th box, $i = 1, 2, ... \aleph$, are sufficiently large. The number of boxes, $\aleph$, is assumed to be also large to decrease statistical errors from the discretization of functions of the excitation energy $U$. Therefore, the statistical conditions are properly accounted for. Then, the average level density $\rho_i$ in the point $i$ was calculated as $\rho_i = L_i / U_{\text{box}}$. The box size $U_{\text{box}}$ plays the same role as the averaging Gauss width parameter $\tilde{\Gamma}$ in Strutinsky's SCM [47]. However, it is more convenient to use approximately the sample number $\aleph$ as the averaging parameter to check the so-called plateau condition. The plateau condition of insensitivity of the value of $K$ was mainly checked for $\aleph = 5-8$ in our mean level density calculation of dots from the experimental spectra [69]. Thus, as in the SCM [47], the result of calculations of the averaged density $\rho_i$, $y_i^{\exp}$ in Eq. (39), is almost independent of the averaging parameter $\aleph$, as in the SCM. The dot bars in Fig. 4 show the statistical part of the errors, related to the distribution of quantum states over sample boxes, $\Delta y_i = 1/\sqrt{L_i}$ [7]. Notice also that the results of our LMSF calculations are almost insensitive to the weights, related to a small $\Delta y_i$, because they are much smaller than the values of $y_i$. The results of the calculations of dots are also almost independent of the small error bars in the horizontal energy direction. Their dependence on the discretization is reduced to the minimum taking $U_i$, as usual, at the mean weighted value within the sample.

In Fig. 4 the MMA approach, Eq. (10), in terms of the specific approximations "2a", Eq. (12), "2b", Eq. (13), and "1", Eq. (11), at the minimum value of $\sigma$ (Table 1) is compared with the standard Fermi gas (FG) formula (17) versus experimental data. For the relative shell corrections in $^{150}$Sm (a), $^{166}$Ho (b), and $^{240}$Pu (c), $\mathcal{E}_{\text{sp}}$ is taken from Ref. [61] (see Table 1), where the shell corrections are calculated within the pairing (HFB) formalism. The binding energy is equal approximately to the ETF energy. Fig. 4 shows a nice agreement of the MMA results for the level density $\rho$ with those of experimental data for several deformed nuclei at low excitation energies.

The pairing effects are not shown for these nuclei because they are small as compared with the typical condensation energies $E_{\rm cond} \sim 1-2$ MeV, see Section 4. The reason is that there are a lot of levels at very small excitation energies which are much smaller than this typical estimate of $E_{\rm cond}$ in the spectra given by Ref. [69]. For instance, the nucleus $^{166}$Ho [Fig. 4(b) and Fig. 5(b)] has a lot of excited states in a very low-energy range which is significantly smaller than 1 MeV. Indeed, one finds here about 100 levels and even much more quantum states when accounting for their $2I_i+1$ degeneracies.

Shell effects of the MMA approach are clearly important for nuclei $^{166}$Ho [perfect MMA2b approach in Fig. 4(b)], and $^{240}$Pu [MMA2a in Fig. 4(c)], see also Table 1. However, it is difficult to determine the shell contribution for the nucleus $^{150}$Sm, for which small $\sigma$ for the MMA1 is close to that of the MMA2a approach because of small $\mathcal{E}_{\rm sp}$ depending on the shell and pairing contributions in the numerical results [61]: these contributions cannot be presented separately.

Fig. 5 presents similar plots to those in Fig. 4 but for low energy states in $^{144}$Sm (a), $^{166}$Ho (b), $^{208}$Pb (c) and $^{230}$Th (d), within different MMA approximations by the same condition of the smallest LMSF parameter $\sigma$, Eq. (39) (we keep the same nucleus $^{166}$Ho for convenience of comparison). Several realistic values of relative energy shell (including pairing) corrections are given in Table 1. In the spherical semi-magic $^{144}$Sm (a) and magic $^{208}$Pb (c) nuclei, for which one has typical condensation energies 1-2 MeV (see Table 1), the smallest $\sigma$ were found for the MMA2 approaches (heavy dashed lines). In addition to the notations of Fig. 4, in the same two panels (a) and (c), the rare dashed lines, marked by primes, test shifts of the excitation energies $U_{\rm eff}$, Eq. (36), by condensation energies $E_{\rm cond}$, Eq. (37). Thus, we take into account the pairing correlations; see Table 1 and arrows in Fig. 5 (a,c). As can be seen from Fig. 5 (a,c), the comparison between the results of the MMA approximation [MMA2b in (a) and MMA2a in (c) with minimal $\sigma$ (Table 1)] and experimental data are significantly improved for spherical nuclei $^{144}$Sm (a) and $^{208}$Pb (c) by taking into account the pairing condensation energies (cf. solid and dashed lines). According to these plots and Table 1, the level distributions in these two nuclei is completely different as compared, e.g., with those of the nucleus $^{166}$Ho [(b) in Figs. 4 or 5]. In the nuclei $^{144}$Sm and $^{208}$Pb, there are no levels below the value of $E_{\rm cond} \sim 1$ MeV, but a lot of levels, and even much more excitation states above this value. We should point out, however, that the experimental observation of pairing effect in these two nuclei is not easy because the pairing collapse (red arrows) appears near the first dot in Fig. 5. Notice also that the pairing effect is seen more pronouncedly from the shell structure MMA2b and MMA2a approximations in $^{144}$Sm (a) and $^{208}$Pb (c), respectively, even a little bit clearly in the plot (c) [cf. solid (without) and dashed (with prime) red lines]. The alert reader could note the difference in these results from those of Ref. [48] because of the importance of taking into account here the isotopic asymmetry as compared to the calculations in Ref. [48] for isotopically symmetric case. By reasons mentioned above, there is almost no pairing effects in the right plots (b) for $^{166}$Ho and (d) for $^{230}$Th, both nuclei are very deformed, especially $^{166}$Ho. However, the shell effects in the level density $\rho(E,N,Z)$ within the MMA2b and MMA2a approaches are more pronounced, see Table 1. The values of $K$ for spherical semi-magic $^{144}$Sm (a), $K \approx 34$ MeV, and magic $^{208}$Pb (c), $K = 24$ MeV, nuclei are significantly larger than those for $^{166}$Ho (b), $K \approx 13$ MeV, and for $^{230}$Th, $K \approx 11$ MeV. Note also that for $^{208}$Pb one has a larger value than that found from the theoretical results shown in Fig. 3 but smaller than for model-dependent experimental results [18], valid mostly for neutron resonances.

Table 1. Inverse level density parameter $K$ (sixth) and its error (seventh column), found by the LMSF with the relative accuracy $\sigma$ (eight column), for the nuclei shown in Figs. 4, 5, 6; see the first column (the asterisk superscript shows the superfluid nuclei with a finite condensation energy), and Eqs. (11), (12), (13) and (17). The second column presents the MMA and FG approaches. The third column places the unified relative shell and pairing corrections $\mathcal{E}_{sp}$, Eq. (9) (from Ref. [61]). The fourth and fifth columns show the condensation energies $E_{cond}$, Eq. (37), and total excitation energies $U_c^{tot}$ for the phase transition from the superfluid to normal nuclear liquid phases, Eq. (34) in case of notable pairing (marked by asterisks), respectively.

| Nuclei | MMA | $\mathcal{E}_{sp}$ | $E_{cond}$ (MeV) | $U_c^{tot}$ (MeV) | $K$ (MeV) | $\Delta K$ (MeV) | $\sigma$ |
|---|---|---|---|---|---|---|---|
| $^{140}$Nd* | 2b | 0.14 | 0.72 | 2.25 | 30.3 | 1.4 | 2.7 |
|  | 2a |  | 1.8 | 5.5 | 12.4 | 1.6 | 6.9 |
|  | 1 |  | 1.6 | 4.9 | 13.8 | 2.7 | 12.8 |
|  | FG |  | 0 |  | 19.1 | 0.5 | 2.3 |
| $^{142}$Nd* | 2b | 0.44 | 0.69 | 2.2 | 31.5 | 1.3 | 2.2 |
|  | 2a |  | 1.8 | 5.7 | 11.9 | 1.6 | 8.0 |
|  | 1 |  | 1.6 | 5.0 | 13.5 | 2.3 | 9.3 |
|  | FG |  | 0 |  | 19.8 | 0.4 | 1.3 |
| $^{144}$Sm* | 2b | 0.37 | 0.64 | 2.0 | 34.2 | 1.2 | 1.4 |
|  | 2a |  | 1.7 | 5.3 | 12.8 | 1.7 | 5.7 |
|  | 1 |  | 1.5 | 4.7 | 14.6 | 2.5 | 7.0 |
|  | FG |  | 0 |  | 20.8 | 0.4 | 1.1 |
| $^{145}$Nd | 2b | 0.25 | 0 |  | 24.0 | 0.7 | 1.7 |
|  | 2a |  | 0 |  | 11.1 | 0.4 | 2.7 |
|  | 1 |  | 0 |  | 9.9 | 0.6 | 4.1 |
|  | FG |  | 0 |  | 10.5 | 0.5 | 2.9 |
| $^{150}$Sm | 2b | 0.18 | 0 |  | 27.4 | 1.5 | 3.6 |
|  | 2a |  | 0 |  | 13.6 | 0.3 | 1.7 |
|  | 1 |  | 0 |  | 12.4 | 0.2 | 1.3 |
|  | FG |  | 0 |  | 13.0 | 0.3 | 1.4 |

| Nuclei | MMA | $\mathcal{E}_{sp}$ | $E_{cond}$ (MeV) | $U_c^{tot}$ (MeV) | $K$ (MeV) | $\Delta K$ (MeV) | $\sigma$ |
|---|---|---|---|---|---|---|---|
| $^{166}$Ho | 2b | 0.49 | 0 | | 12.8 | 0.3 | 2.3 |
| | 2a | | 0 | | 5.5 | 0.3 | 8.6 |
| | 1 | | 0 | | 4.5 | 0.4 | 13.1 |
| | FG | | 0 | | 4.7 | 0.4 | 11.4 |
| $^{208}$Pb$^*$ | 2b | 1.77 | 0.36 | 1.1 | 60.3 | 3.4 | 4.2 |
| | 2a | | 0.92 | 2.9 | 23.9 | 0.9 | 3.2 |
| | 1 | | 0.69 | 2.1 | 31.9 | 1.4 | 3.3 |
| | FG | | 0 | | 38.3 | 1.6 | 3.6 |
| $^{230}$Th | 2b | 0.55 | 0 | | 27.4 | 1.5 | 3.4 |
| | 2a | | 0 | | 11.3 | 0.3 | 2.2 |
| | 1 | | 0 | | 10.2 | 0.4 | 3.2 |
| | FG | | 0 | | 10.8 | 0.4 | 2.4 |
| $^{240}$Pu | 2b | 0.66 | 0 | | 30.6 | 2.2 | 3.8 |
| | 2a | | 0 | | 12.1 | 0.5 | 3.2 |
| | 1 | | 0 | | 11.3 | 0.7 | 4.1 |
| | FG | | 0 | | 12.0 | 0.6 | 3.2 |

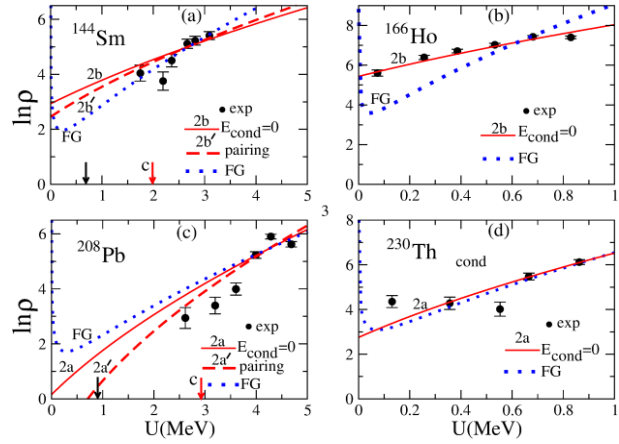

Fig. 5. The same as in Fig. 4 but for a few other nuclei to show the pairing effect in magic and semi-magic nuclei [see heavy rare dashed lines marked by primes in panels (a) and (c)]. Arrows show the pairing condensation energy $E_{cond}$ [left black, Eq. (37)] and the total critical excitation energy $U_c^{tot}$ [right red, Eq. (34)]. Experimental data (dots) are obtained by the same sample method from the spectrum of the database [69].

Fig. 6 shows the level density $\rho(E,N,Z)$ for three other remarkable nuclei with significant differences in pairing and shell effects. As in Fig. 5, we show results for the nucleus $^{145}$Nd (c) with a large number of the low energy states below excitation energy of about 1.5 MeV. For $^{140}$Nd (a) and $^{142}$Nd (b), one has a very small number of low energy states below almost the same energy (see ENSDF database [69] and Table 1). But there are many states in the nuclei $^{140,142}$Nd with excitation energies of above approximately 1.5 MeV, up to essentially larger excitation energy, with definite energies and spins, and in this sense, one can again consider these excitation spectra as almost complete. According to Ref. [61] and Table 1, the shell and pairing effects, measured by $\mathcal{E}_{sp}$, are significant in these nuclei, larger in the spherical $^{142}$Nd than in the slightly deformed isotopes $^{140}$Nd and $^{145}$Nd.

In Fig. 6 and Table I, the results of the MMA1 and MMA2 approximations, Eqs. (11) and (12) [Eq. (13)], respectively, are compared with the FG approach, Eq. (17), and with experimental data. The results of the MMA2 approaches, Eqs. (12) and (13), as the dominating shell effect case (ii) [$\xi \gg 1$, Eq. (9)] modified by pairing correlations with realistic relative shell correction, $\mathcal{E}_{sp}$ (Ref. [61]), are shown versus those of a small such effects, the MMA1 (i) and FG approximations (Table 1), valid at $\xi \ll 1$. The results of the limit of the dominating MMA2 approach to a very small value of $\mathcal{E}_{sp}$, but still within the shell structure case (ii), Eq. (13), named as the MMA2b approach, are also shown in Fig. 6 and Table 1; see "2b" and "2b'" lines, neglecting and taking into account pairing effects, respectively. They are often in clear contrast to the results of the MMA1 approximation [Eq. (11)], and even more with those of the FG asymptotical full saddle-point approach, Eq. (17); see also Table 1. The best LMS fit, with the smallest $\sigma$, for the low excitation-energy spectrum in the semi-magic nucleus $^{142}$Nd [Fig. 6(b)] is achieved for the MMA2b (see Table 1) approach, especially well for the pairing modified 2b approximation, at the inverse level density parameter $K \approx 32$ MeV (see also Table 1). The critical value of the excitation energy, $U_c^{tot}$ (red arrow), for this nucleus is well distinguished from the beginning of the excitation energy by a distance of the order of the condensation energy $E_{cond}$ (black arrow), which more clearly observed in $^{140}$Nd than in the cases of the semi-magic and magic nuclei, $^{144}$Sm and $^{208}$Pb, respectively. For the even-odd nucleus $^{145}$Nd [Fig. 6(c)], there is no notable contribution from pairing correlations. The value of $K$ for $^{145}$Nd, $K \approx 24$ MeV, is smaller than that for $^{140,142}$Nd (see Table 1), in good agreement with the results for a similar odd-odd nucleus $^{166}$Ho.

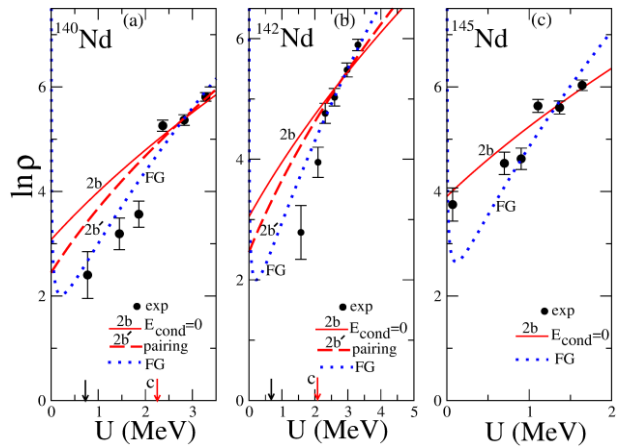

Fig. 6. Level densities, $\ln\rho(E,N,Z)$, for low energy states in nuclei $^{140}$Nd (a) $^{142}$Nd (b), and $^{145}$Nd (c) were calculated within different approximations. The MMA2b approximations (with and without pairing) are shown by the red solid "2b" [Eq. (13) with $E_{cond}=0$] and heavy dashed "2b'" [Eq. (13) modified in (a) by pairing shift with Eq. (37) for the condensation energy $E_{cond}$ (Table 1)] lines. The Fermi Gas approach, Eq. (17), is presented by the rare blue dotted lines (for chemical potentials one assumes $\mu_n \approx \mu_p \approx \mu$, where μ = 40 MeV). Experimental dots with error bars are obtained from the ENSDF database [69] by using the sample method [11,51], as explained in the text.

In Fig. 6 (a) for $^{140}$Nd the line "2b'" is located closer to experimental data than the line "2b". For the nuclei $^{142}$Nd [Fig. 6(b)] and $^{145}$Nd [Fig. 6(c)] the MMA2b results, Eq. (13), are also significantly better in agreement with the experimental data as compared to the results of MMA1 and MMA2a approximations (see Table 1 for the same nuclei). For these nuclei, the MMA1 [Eq. (11)] approximation is characterized by larger $\sigma$ (see Table 1). In contrast to the case of $^{145}$Nd [Fig. 6(c)] with excitation energy spectrum having a large number of low energy states below about 1.5 MeV, for the isotopes $^{140,142}$Nd [Fig. 6(a,b)] one finds almost no such states in the same energy range. As result, one obtains a significant pairing effect in a semi-magic nucleus $^{142}$Nd and even-even isotope $^{140}$Nd as compared to the even-odd nonmagic nucleus $^{145}$Nd. However, for the nuclei $^{140,142}$Nd, the Fermi gas results better agree with data than those of the MMA2b ones, even for the line "2b'" with the pairing effect included. Note that in these nuclei the values of $U_{eff}$, which appear in the LMSF procedure, are small. In this case a more accurate calculation with a piecewise smooth approximation for the modified Bessel function using the exact zero asymptote for $U_{eff}$ improves the results.

The MMA at low excitation energies clearly manifests an advantage over the standard Fermi gas asymptote (FG) for most of nuclei because of no divergences of the MMA in the limit to small excitation energies $U$, $U \to 0$. This is clearly seen also analytically, in particular in the FG limit, Eq. (17); see the general asymptotic expression (14). It is, obviously, in contrast to any MMAs combinatorics expressions (15) in this limit; see also Eqs. (11)-(13). The MMA1 results are sometimes close to those of the FG approach for some considered nuclei ($^{140}$Nd, see Table 1). The reason is that their differences are essential only for extremely small excitation energies $U$, where the MMA1 approach is finite while other, FG and SFG [Eqs. (17) and (16)] approaches are divergent. However, sometimes there are almost no experimental data for excited states in the range of their differences.

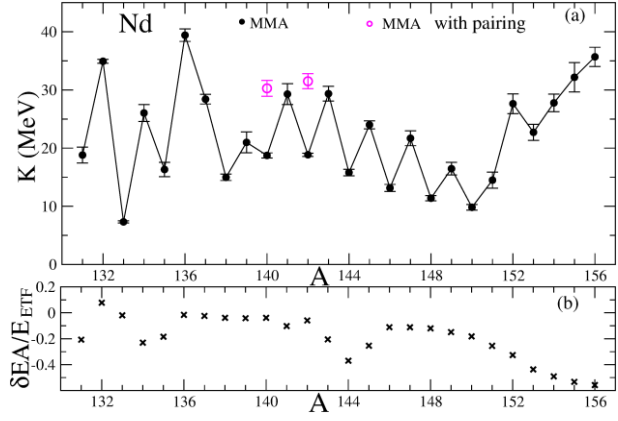

Fig. 7. The inverse level-density parameter $K$ as a function of the particle number $A$ [Fig. 7(a)], and the energy shell corrections $\delta E$ in units of the background energy ($E_{BG} \approx E_{ETF}$) per particle [Fig. 7(b)] for the long isotope Nd chain (from Refs. [61] and [51]). The MMA approximations are taken at the minimal $\sigma$ (mostly MMA2b). Experimental data (closed dots) are obtained by the sample method from the spectrum of the database [69]. Pairing condensation effects are shown by open dots.

Fig. 7 presents the inverse level-density parameters $K$ (with errors) as function of the particle numbers $A$ for a long chain of the Nd isotopes, $A = 131-156$. These experimental data may be incomplete. Nevertheless, it might be helpful to present a comparison between theory and experiment to check general common effects of the statistical isotopic asymmetry, shell and pairing properties in a wider range of nuclei around the $\beta$-stability line. In panel (a) of this figure, the close dots for the MMAs are those taken with the smallest standard relative critical-error LMSF parameter $\sigma$ [Eq. (39)] among all MMA approximations, Eqs. (11)-(13) (mostly the MMA2b results are presented). In panel (b), the relative shell and pairing correction energies $\mathcal{E}_{sp}$, Eq. (9), are also presented, taken approximately in units of the smooth binding energy, $E_{BG} \approx E_{ETF}$, per particle; see Ref. [61]. We obtained values of $K$ for the low-energy states range which are essentially different from the results for neutron resonances. Indeed, our results for $K$ are significantly larger than those for the neutron resonances. As seen from Fig. 7, one finds a saw-toothed behavior of $K(A)$ (low values for even-even and high ones for even-odd nuclei) as a function of the particle numbers $A$ with a remarkable shell oscillation (cf. Fig. 7(a) with its (b) panel). We may conclude from this Figure that the shell effects are quite important and should be taken into account in our statistical calculations. Pairing effects in the two even-even nuclei $^{140}$Nd and $^{142}$Nd (shown by open dots) significantly increase the values of $K$; see Table 1. Thus, the pairing correlations lead to a smoother behavior of the inverse level density parameter $K$.

As results, the statistically averaged level densities $\rho(E,N,Z)$ for the MMA approximation with a minimal value of the control-error parameter $\sigma$, Eq. (39), in plots of Figs. 4-7 agree well with those of the experimental data with the unique free but physical parameter $K$ of the LMSF procedure. The results of the MMA, and FG approaches for the level densities $\rho(E,N,Z)$ in these figures, and for $K$ in Table 1, do not depend on the cut-off spin factor and moment of inertia because of the summations (integrations) over all spins projections, or over spins, indeed, with accounting for the degeneracy $2I+1$ factor. We do not use empiric free fitting parameters in our calculations, in particular, for the FG results shown in Table 1. This is in contrast, e.g., to model-dependent approaches [18-21], where the level density integrated over spins depends on a cut-off (the moment of inertia times temperature) parameter.

## 6. Conclusions and perspectives

The level density $\rho(U)$ as a function of the excitation energy $U$ is obtained analytically within the micro-macroscopic approach (MMA) in the semiclassical periodic-orbit approximation accounting for the nuclear shell structure, pairing correlations, and isotopic asymmetry. The analytical expression for the level density has both the correct asymptotes for large (Fermi gas) and small (combinatoric) excitation energies. Experimental data for the averaged level density were obtained by the Least Mean Square Fitting (LMSF) with one free physical parameter – inverse level density parameter $K$ – by the sample method. A few other constants having typical values were used but never changed in all these calculations and reported in the text in order to reproduce our results. The results of the LMSF are almost independent of the number of sample boxes $\aleph$ (or energy interval $U_{\text{box}}$). It plays a similar role as obeying a plateau condition with respect to the stable values of $K$, as the Gaussian width and correction polynomial degree in Strutinsky's smoothing procedure. Therefore, our results for empiric coarse-graining level density at states of low excitation energies are weakly dependent on the model. They were obtained from the available experimental data with a reasonable completeness on states with low excitation energies accounting for the spin degeneracy and satisfying the required statistical properties. Reasonable agreement between the inverse level density parameter, $K = A/a$, as a function of the particle number $A$ and the experimental data for the neutron resonances was obtained for nuclei with intermediate particle numbers $A \approx 100-150$. The spin-orbit interaction is effectively taken into account phenomenologically by shifting the curve $K = K(A)$ in about half of the period of the major shell structure. For the coarse-graining level density, one obtains basic properties, namely, the shell, paring, and asymmetry effects for several heavy spherical and deformed nuclei for particle numbers $A \approx 100-200$. As shown, taking many typical nuclei, these effects significantly influence the statistically averaged level density. The results for the MMA level density $\rho(E,N,Z)$, which were integrated over all nuclear spins and their projections, are beyond the small-spin approximation, and can be applied for the low excitation energies below neutron resonances. We clarified the phenomenological Back Shifted Fermi Gas model as the effect of the corrections in the MMA expansion over inverse powers of the entropy for a large entropy. The results of the Effective Temperature Constant Model is explained by combinatoric expansion of the MMA level density for small entropy (or excitation energy). The MMA approximation takes into account the interparticle interaction through the extended Thomas-Fermi (ETF) counterparts of the periodic orbit theory (POT). The POT reproduces well the shell effects in the level density through the grand-canonical ensemble potential for low excitation energies. For even-even nuclei with shifted spectra of excitation states in a finite interval larger or of the order of the condensation energy $E_{cond} \approx 1-2$ MeV we evaluated the pairing effects through the shifts $E_{cond}$ for the excitation energy $U$. In several semi-magic and magic nuclei we estimated the critical excitation energy for the pairing collapse and explained the difficulties of the observation of the phase transition from a superfluid to the normal nuclear state in semi-magic and magic nuclei. Consistent statistical averaging of the level density was emphasized in our MMA calculations.

For perspectives, we suggest using our results for the statistical analysis of the collective quantum spectra of deformed rotating nuclei obtained, e.g., in the two-neutron transfer (t,p) reactions. We are planning to calculate fission widths within the Bohr&Wheeler theory by using the coarse-graining MMA level density, regular at small excitation energies. As the semiclassical POT approximation is the better the larger the particle numbers, our MMA approach could be applied for metallic clusters and quantum dots, and also in nuclear astrophysics.

The authors gratefully acknowledge D. Bucurescu, R.K. Bhaduri, M. Brack, C.M. Ko, A.N. Gorbachenko, J.B. Natowitz, E. Pollacco, V.A. Plujko, P. Ring, and A. Volya for creative discussions. This work was supported in part by the budget program "Support for the development of priority areas of scientific research", the project of the Academy of Sciences of Ukraine (Code 6541230, no. 0122U000848). S. Shlomo and A.G. Magner are partially supported by the US Department of Energy under Grant no. DE-FG03-93ER-40773. U.V. Grygoriev is partially supported by Van Swinderen Institute, Faculty of Science and Engineering of the University of Groningen in the Netherlands.

REFERENCES

1. H. Bethe. An attempt to calculate the number of energy levels of a heavy nucleus. Phys. Rev. 50 (1936) 332.

2.  T. Ericson. The statistical model and nuclear level densities. Adv. Phys. 9 (1960) 425.
3.  A. Gilbert and A.G.W. Cameron. A composite nuclear-level density formula with shell corrections. Can. J. Phys. 43 (1965) 1446.
4.  Aa. Bohr, B.R. Mottelson. *Nuclear structure.* Vol. 1 (Benjamin, New York, 1967).
5.  V.S. Stavinsky. Nuclear level density. Sov. J. Part. Nucl. 3 (1972) 417.
6.  A.V. Ignatuyk, G.N. Smirenkin, A.S. Tishin. Phenomenological description of energy dependence of the level density parameter. Sov. J. Nucl. Phys. 21 (1975) 255.
7.  L.D. Landau, E.M. Lifshitz. *Statistical physics.* Part 1 (Oxford: Pergamon Press, 1980) 544 p.
8.  S.K. Kataria, V.S. Ramamurthy, S.S. Kapoor. Semiempirical nuclear level density formula with shell effects. Phys. Rev. C 18 (1978) 549.
9.  A.V. Ignatyuk. *Statistical Properties of Excited Atomic Nuclei* (Moscow: Energoatomizdat, 1983) (Rus)
10. M.K. Grossjean, H. Feldmeier. Level density of a Fermi gas with pairing interactions. Nucl. Phys. A 444 (1985) 113.
11. Yu.V. Sokolov. *Level Density of Atomic Nuclei* (Moscow: Energoatomizdat, 1990) (Rus)
12. S. Shlomo, J.B. Natowitz. Level density parameter in hot nuclei, Phys. Lett. B 252 (1990) 187.
13. S. Shlomo, J.B. Natowitz. Temperature and mass dependence of level density parameter. Phys. Rev. C 44 (1991) 2878.
14. S. Shlomo. Energy level density of nuclei. Nucl. Phys. A 539 (1992) 17.
15. S. Goriely. A new nuclear level density formula including shell and pairing correction in the light of a microscopic model calculation. Nucl. Phys. A 605 (1996) 28.
16. P. Demetriou, S. Goriely. Microscopic nuclear level densities for practical applications. Nucl. Phys. A 695 (2001) 95.
17. Y. Alhassid, G.F. Bertsch, L. Fang. Nuclear level statistics: Extending shell model theory to higher temperatures. Phys. Rev. C 68 (2003) 044322.
18. T. von Egidy, D. Bucurescu. Systematics of nuclear level density parameters. Phys. Rev. C 72 (2005) 044311.
19. T. von Egidy, D. Bucurescu. Spin distribution in low-energy nuclear level schemes. Phys. Rev. C 78 (2008) 051301(R).
20. N.U.H. Syed et al. Level density and γ-decay properties of closed shell Pb nuclei. Phys. Rev. C 79 (2009) 024316.
21. T. von Egidy, D. Bucurescu. Experimental energy-dependent nuclear spin distributions. Phys. Rev. C 80 (2009) 054310.
22. Y. Alhassid et al. Direct microscopic calculation of nuclear level densities in the shell model Monte Carlo approach. Phys. Rev. C 92 (2015) 024307.
23. Y. Alhassid et al. Benchmarking mean-field approximations to level densities. Phys. Rev. C 93 (2016) 044320.
24. R. Sen'kov, V. Zelevinsky. Nuclear level density: Shell-model approach. Phys. Rev. C 93 (2016) 064304.

25. S. Karampagia, V. Zelevinsky. Nuclear shape transitions, level density, and underlying interactions. Phys. Rev. C 94 (2016) 014321.
26. A. Heusler et al. Complete identification of states in $^{208}$Pb below $E_x = 6.2$ MeV. Phys. Rev. C 93 (2016) 054321.
27. V. Zelevinsky, S. Karampagia. Nuclear level density and related physics. EPJ Web Conf. 194 (2018) 01001.
28. V. Zelevinsky, M. Horoi. Nuclear level density, thermalization, chaos, and collectivity. Prog. Part. Nucl. Phys. 105 (2019) 180.
29. S. Karampagia, V. Zelevinsky. Nuclear shell model and level density. Int. J. Mod. Phys. E 29 (2020) 2030005.
30. P. Fanto, Y. Alhassid. State densities of heavy nuclei in the static-path plus random-phase approximation. Phys. Rev. C 103 (2021) 064310.
31. V.M. Kolomietz, A.G. Magner, V.M. Strutinsky. Shell effects in rotating nuclei. Sov. J. Nucl. Phys. 29 (1979) 758.
32. V.A. Plujko, O.M. Gorbachenko. Effect of vibrational states on nuclear level density. Phys. Atom. Nucl. 70 (2007) 1643.
33. B.K Jennings, R.K. Bhaduri, M. Brack. Semiclassical approximation in a realistic one-body potential. Nucl. Phys. A 253 (1975) 29.
34. M. Brack, B.K. Jennings, Y.H. Chu. On the extended Thomas-Fermi approximation to the kinetic energy density. Phys. Lett. B 65 (1976) 1.
35. W.E. Ormand. Estimating the nuclear level density with the Monte Carlo shell model. Phys. Rev. C 56 (1997) R1678.
36. M. Gutzwiller. Periodic orbits and classical quantization conditions. J. Math. Phys. 12 (1971) 343.
37. M. Gutzwiller. *Chaos in Classical and Quantum Mechanics* (New York: Springer-Verlag, 1990).
38. R. Balian, C. Bloch. Distribution of eigenfrequencies for the wave equation in a finite domain: III. Eigenfrequency density oscillations. Ann. Phys. 69 (1972) 76.
39. V.M. Strutinsky, A.G. Magner. Quasiclassical theory of nuclear shell structure. Sov. J. Part. Nucl. 7 (1976) 138.
40. A.G. Magner, V.M. Kolomietz, V.M. Strutinsky. Gross-shell effects in the single-particle level distribution with fixed angular momentum projection. Sov. J. Nucl. Phys. 28 (1978) 764.
41. M. Brack, R.K. Bhaduri. *Semiclassical Physics*, Frontiers in Physics. Vol. 96 (Boulder: Westview Press, 2003).
42. A.G. Magner et al. Shell structure and orbit bifurcations in finite fermion systems. Phys. Atom. Nucl. 74 (2011) 1445.
43. M. Brack, C. Guet, H.-B. Håkansson. Selfconsistent semiclassical description of average nuclear properties – a link between microscopic and macroscopic models. Phys. Rep. 123 (1985) 275.
44. V.M. Kolomietz, A.I. Sanzhur, S. Shlomo. Self-consistent mean-field approach to the statistical level

density in spherical nuclei. Phys. Rev. C 97 (2018) 064302.
45. V.M. Kolomietz and S. Shlomo. *Mean Field Theory* (Singapore: World Scientific, 2020) 565 p.
46. V.M. Strutinsky. Shell effects in nuclear masses and deformation energies. Nucl. Phys. A 95 (1967) 420; "Shells" in deformed nuclei. Nucl. Phys. A 122 (1968) 1.
47. M. Brack et al. Funny hills: The shell-correction approach to nuclear shell effects and its applications to the fission process. Rev. Mod. Phys. 44 (1972) 320.
48. A.G. Magner et al. Semiclassical shell-structure micro-macroscopic approach for the level density. Phys. Rev. C 104 (2021) 044319.
49. A.G. Magner et al. Shell-structure and asymmetry effects in level densities. Int. J. Mod. Phys. E 30 (2021) 2150092.
50. A.G. Magner et al. Level density within a micro-macroscopic approach. Nucl. Phys. A 1021 (2022) 122423.
51. A.G. Magner et al. Microscopic-macroscopic level densities for low excitation energies. Low Temp. Phys. 48 (2022) 920.
52. A.I. Levon et al. High-resolution study of excited states in $^{158}$Gd with the $(p,t)$ reaction. Phys. Rev. C 102 (2020) 014308.
53. N. Bohr, J.A. Wheeler. The mechanism of nuclear fission. Phys. Rev. 56 (1939) 426.
54. V.M. Strutinsky. On the nuclear level density in case of an energy gap. Proc. Int. Conf. on Nucl. Phys. (Paris, 1958) p. 617.
55. A. Bohr, B.R. Mottelson, D. Pines. Possible analogy between the excitation spectra of nuclei and those of the superconducting metallic state. Phys. Rev. 110 (1958) 936.
56. S.T. Belyaev. Effect of pairing correlations on nuclear properties. Mat. Fys. Medd. Dan. Vid. Selsk. 31 (1959) 3.
57. J. Bardeen, L.N. Cooper, J.R. Schrieffer. Theory of superconductivity. Phys. Rev. 108 (1957) 1175.
58. P. Ring, P. Schuck. The Nuclear Many-Body Problem (New York: Springer-Verlag, 1980) 716 p.
59. N.N. Bogoliubov. A new method in the theory of superconductivity. I. Sov. Phys. JETP. 7 (1958) 41.
60. A. Sedrakian, J.W. Clark. Superfluidity in nuclear systems and neutron stars. Eur. Phys. J. A 55 (2019) 167.
61. P. Moeller et al. Nuclear ground-state masses and deformations: FRDM (2012). At. Data Nucl. Data Tables 109-110 (2016) 1.
62. U. Mutz, P.Ring. On the pairing collapse in nuclei at high angular momenta. J. Phys. G 10 (1984) L39.
63. J.L. Edigo et al. On the validity of the mean field approach for the description of pairing collapse in finite nuclei. Phys. Lett. B 154 (1985) 1.
64. V.M. Strutinsky et al. Semiclassical interpretation of the gross-shell structure in deformed nuclei. Z. Phys. A 283 (1977) 269.
65. A.V. Ignatyuk, Yu.V. Sokolov. Density of "particle-hole" excited states in shell model. Yad. Fiz. 16 (1972) 277.

66. W. Dilg et al. Level density parameters for the back-shifted fermi gas model in the mass range $40 < A < 250$. Nucl. Phys. A 217 (1973) 269.
67. B.K. Agrawal, S. Shlomo, V.K. Au. Determination of the parameters of a Skyrme type effective interaction using the simulated annealing approach. Phys. Rev. C 72 (2005) 014310.
68. D.V. Gorpinchenko, A.G. Magner, J. Bartel. Semiclassical and quantum shell-structure calculations of the moment of inertia. Int. J. Mod. Phys. E 30 (2021) 2150008.
69. National Nuclear Data Center On-Line Data Service for the ENSDF (Evaluated Nuclear Structure Data File) database, http://www.nndc.bnl.gov/ensdf.